%% file: OPTICAmeetings.tex
\documentclass[letterpaper,10pt]{article} 

\usepackage{opticameet3} 

\newcommand\authormark[1]{\textsuperscript{#1}}

\usepackage{amsmath,amssymb}
\usepackage{pgfplots} 
\usepackage{adjustbox}

\usepackage[colorlinks=true,bookmarks=false,citecolor=blue,urlcolor=blue]{hyperref} 
\pgfplotsset{compat=1.16}
\usepgfplotslibrary{colorbrewer}
\begin{document}

\title{Dual optical frequency comb downconversion of D-band mm-wave signals}


\author{Callum Deakin,\authormark{1,*} Zichuan Zhou,\authormark{2} Ronit Sohanpal,\authormark{2} and Zhixin Liu\authormark{2}}

\address{\authormark{1}Nokia Bell Labs, 600 Mountain Ave, Murray Hill, NJ, USA\\
\authormark{2}Department of Electronic and Electrical Engineering, University College London, UK}

\email{\authormark{*}callum.deakin@nokia-bell-labs.com} 


\begin{abstract}
We demonstrate a dual optical frequency comb concept that down-converts arbitrary narrowband D-band (110-170 GHz) signals to baseband without any filter or optical/RF frequency tuning, using low frequency RF components.
\end{abstract}
{\vspace{0.1cm}}
\section{Introduction}

D-band millimetre-wave frequencies (110-170 GHz) have received extensive research interest due to broad unlicensed frequency availability, low atmospheric absorption, and the potential to dramatically expand the capacity of cellular and backhaul wireless links, particularly in dense urban environments~\cite{maiwald2023review}. Operating at such high frequencies is of particular interest in integrated sensing and communications applications (ISAC) due to the increased spatial-temporal resolution~\cite{liu2022integrated}, while the enhanced spatial resolution of D-band radar is also desired in industrial, security and avionics/automotive applications~\cite{appleby2007millimeter}.
 
Many efforts have examined photonic-based approaches to microwave signal reception, which involves converting signals into the optical domain, 
avoiding the need for inefficient and power-hungry high frequency electronic components and with the potential to enhance signal quality. However, photonic-based approaches using either independent lasers or frequency combs require precise tuning of the laser frequency, and are generally susceptible to laser phase noise. Mitigating these requires either very stable cavity references that are impractical for integration or complex laser locking schemes~\cite{kassem2024photonic,largo2017frequency}, adding power consumption, cost and size. Recently, dual optical frequency combs have emerged as a technique that allows for downconversion without frequency tuning via compressive sensing~\cite{klee2017dual}, with advantages including insensitivity to laser phase noise~\cite{deakin2020noise} and potential improved sensitivity without needing broadband photodetection. However, previous demonstrations of the dual comb technique were limited to the low frequency ($<40$ GHz) region and relied on optical filter arrays that are complex/impractical for photonic integration~\cite{alshaykh2019rapid,deakin2022frequency,klee2017dual,o2022architecture}.

In this paper, we demonstrate a D-band receiver using phase-coherent dual optical frequency combs in combination with a $>$200-GHz-bandwidth thin-film lithium niobate (TFLN) Mach-Zehnder modulator (MZM). Compared to conventional electronic methods, our method eliminates the need for mm-wave components (e.g. mixer, amplifier, filter). Compared to laser-based microwave photonics methods, we eliminate the complex high-frequency optical phase-locked loop and the need for ultra-low linewidth lasers. To the best of our knowledge, we present the first use of the dual-comb technique for D-band reception that allows for downconversion to baseband across the D-band, without the use of optical frequency tuning or optical filters.

\section{Dual optical frequency comb D-band receiver concept}   
\begin{figure}[b]
\vspace{-0.5cm}
   \centering
    \includegraphics[width=0.55\linewidth]{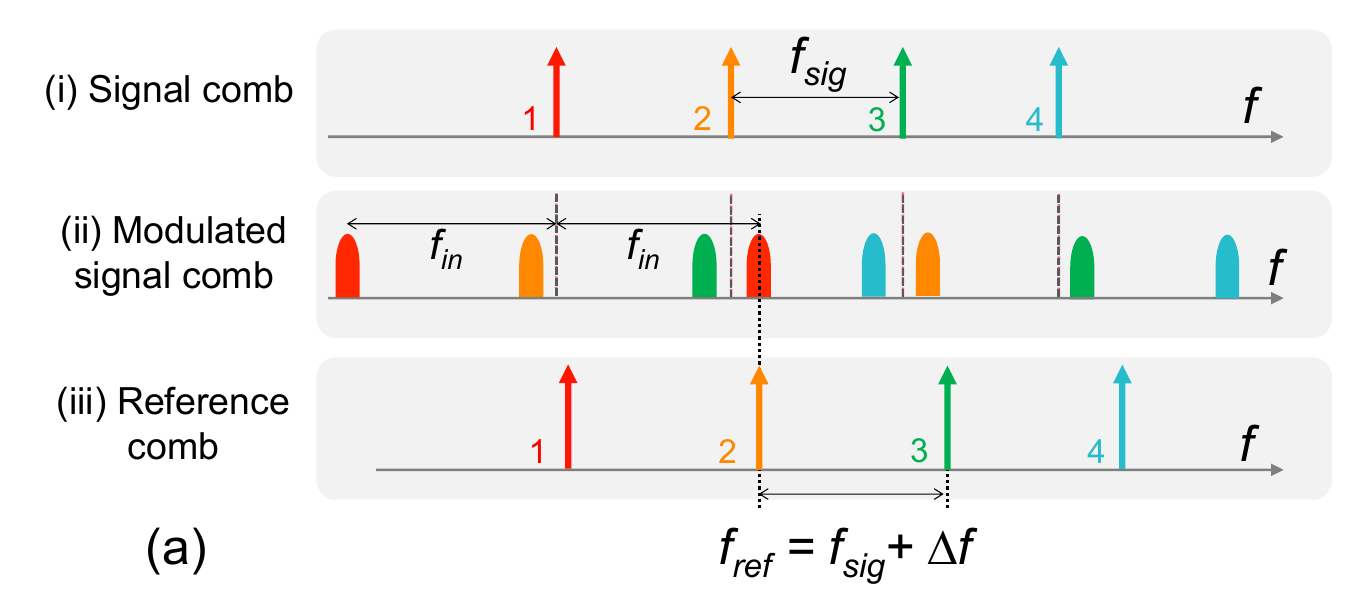}
    \hspace{1cm}
    \input{figs/channel_assignment}
    \vspace{-0.3cm}
    \caption{(a) Dual comb concept. (b) Comb line mapping for detected D-band frequencies 110-170 GHz with $f_{\textnormal{sig}} = 25$~GHz and $f_{\textnormal{ref}} = 27$~GHz. Sig USB and Sig LSB refer to the upper and lower sidebands generated by the MZM respectively.}
    \label{fig:concept}
    \vspace{-0.5cm}

\end{figure}
The dual optical frequency comb technique is based on two frequency combs, here referred to as `signal' and `reference' combs, with offset repetition rates $f_{\textnormal{sig}}$ and $f_{\textnormal{ref}}= f_{\textnormal{sig}} + \Delta f$. The concept is illustrated in Fig.~\ref{fig:concept}(a) for an example of 4 comb lines. Using amplitude modulation, the signal comb with repetition rate $f_{\textnormal{sig}}$ is modulated with the signal of interest, of center carrier frequency $f_{\textnormal{in}}$, resulting in every comb line being modulated and producing the modulated signal at frequencies $\pm f_{\textnormal{in}}$ from every signal comb line, as shown in Fig.~\ref{fig:concept}(a)(ii). When compared to the reference comb, Fig.~\ref{fig:concept}(a)(iii), only the modulated comb line 1 (red) appears at a distance $<\Delta f$ from any single reference comb line (in this case, ref. comb line 2, orange). This signal can then be detected by low speed (bandwidth $\Delta f/2$) balanced photodiodes, due to the beating between the reference comb line 2 (orange) and modulated signal comb line 1 (red). Conversely, all other signal-reference comb line interactions on the photodiodes will be at much higher frequencies and filtered by the low pass response of photodiode-TIA circuit. Specifically, an radio-frequency (RF) signal at frequency $f_{\textnormal{in}}$ will generate a beat signal on the balanced photodiodes of frequency less than $\Delta f$ only for the signal comb line $n_\textnormal{sig}$ and reference comb line $n_\textnormal{ref}$ given by \vspace{-0.2cm}\begin{equation} \label{distortion_index_sig}
    n_\textnormal{sig} = n_\textnormal{ref} \pm (-1)^{\big\lceil \frac{f_{\textnormal{in}}}{2f_\textnormal{sig}}\big\rceil}  \bigg\lceil \frac{f_{\textnormal{in}}}{f_\textnormal{sig}}\bigg\rceil, \hspace{1cm} n_\textnormal{ref} = \Bigg \lceil \frac{\big|f_{\textnormal{in}} - f_{\textnormal{sig}} \big\lceil \frac{f_{\textnormal{in}}}{f_{\textnormal{sig}}} \big\rceil\big|}{\Delta f} \Bigg \rceil
\end{equation}{}where $\lceil x \rceil$ denotes rounding $x$ up to the nearest integer. The $\pm$ in $n_\textnormal{sig}$ represents detection of either the upper or lower sideband resulting from the MZM modulation, respectively. Therefore, the dual comb technique can in theory detect signals at frequencies up to the full optical bandwidth of the frequency combs using only low speed photodiodes, while the maximum bandwidth of the input signal is limited by the comb spacing, $\Delta f$. The equations (\ref{distortion_index_sig}) are plotted in Fig.~\ref{fig:concept}(b) for D-band frequencies 110-170 GHz, using $f_{\textnormal{sig}} = 25$~GHz and $f_{\textnormal{ref}} = 27$~GHz.

\section{Experimental setup}

\begin{figure}[tb]
\vspace{-0.5cm}
   \centering
    \includegraphics[width=0.49\linewidth]{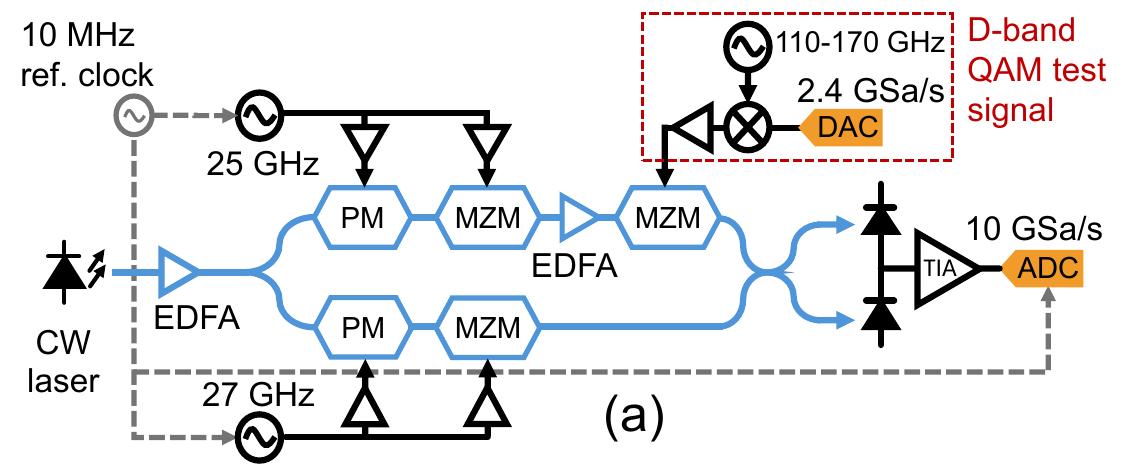}
    \hspace{1cm}
    \input{figs/comb_plot}
        \vspace{-0.3cm}
    \caption{(a) Experimental setup. PM, phase modulator; MZM, Mach-Zehnder modulator; EDFA, erbium doped fiber amplifer; TIA, transimpedance amplifier. (b) Optical spectra for signal comb $f_{\textnormal{sig}} = 25$~GHz, and reference comb $f_{\textnormal{ref}} = 27$~GHz. }
    \label{experiment_setup}
    \vspace{-0.8cm}

\end{figure}

The experimental setup is shown in Fig.~\ref{experiment_setup}(a). We generate two optical frequency combs with a spacing of $f_{\textnormal{sig}} = 25$~GHz and $f_{\textnormal{ref}} = 27$~GHz by modulating a CW laser with cascaded phase and intensity modulators to generate two flat-topped frequency combs that are shown in Fig.~\ref{experiment_setup}(b). In the flat region, the signal comb has optical power per line that varies from 4-9~dBm, while the reference comb varies from 0-5~dBm. The signal comb is then sent to a TFLN-on-silicon MZM biased at null, in which every comb line is modulated with the input RF signal. The 3~dB bandwidth of the MZM is around $100$~GHz but it demonstrates a smooth roll-off in frequency response up to at least 220 GHz, principally defined by the transmission line electrode loss (S21 at 200 GHz $\approx -8$~dB). The coupling losses of the MZM are relatively high (5~dB per coupler) due to inefficient surface grating couplers. As a result, the signal comb is amplified to 23~dBm using an erbium-doped fiber amplifier (EDFA) prior to modulation.

The baseband test signal is a 64-QAM RRC-shaped ($\beta = 0.1$, 101 filter taps) sequence of 7658 symbols generated by a 10-bit 2.4~GSa/s digital to analog converter (DAC). Since a D-band IQ modulator was not available, we use a direct mixer to up-convert the signal to D-band, where the D-band local oscillator is generated by a low phase noise signal generator followed by a $\times 6$ multiplier. The use of a direct mixer (rather than IQ mixer) generates a double side band D-band signal such that the complex conjugate of the base band signal appears on either side of the D-band LO frequency. Although not a strict representation of a typical D-band QAM signal, it generates spectral content that is sufficient to test our concept and is fully indicative of performance: i.e. the 200-MBd double sideband signal is representative of a 400-MBd QAM signal that would be generated by a D-band IQ modulator. The output of the mixer is further amplified by a D-band amplifier before being sent to the MZM via a GSG probe. 

The modulated signal comb and reference comb are then combined in a $2\times2$ 50/50 optical coupler before being sent to a balanced detector. The photodiode-TIA circuit has a bandwidth of 1.6 GHz and a transimpedance gain of approximately $16\times10^3$ V/A. The signal is digitised by a 10-bit 10~GSa/s analog to digital converter (ADC) for offline processing. The downconverted and digitised QAM signals are then processed via standard digital signal processing including a 201-tap least-means-squares (LMS) equaliser embedded with a digital phase locked loop.

\section{Results and discussion}

\begin{figure}[tb]
   \centering
           \input{figs/new_results_fig}
               \vspace{-0.3cm}
    \caption{(a) Optical spectrum excerpt showing a modulated signal comb example (155.82~GHz) and reference comb. Only for $n_\textnormal{ref}= 3$, $n_\textnormal{sig}= -3$ do the combs generate a beat signal on the photodiode. (b) Recovered SNR across D-band frequencies 110-170GHz. (c) Example recovered spectrum for $f_\textnormal{in}=132.252$ GHz and its (d) corresponding constellation.}
    \label{fig:results}
    \vspace{-0.8cm}
\end{figure}
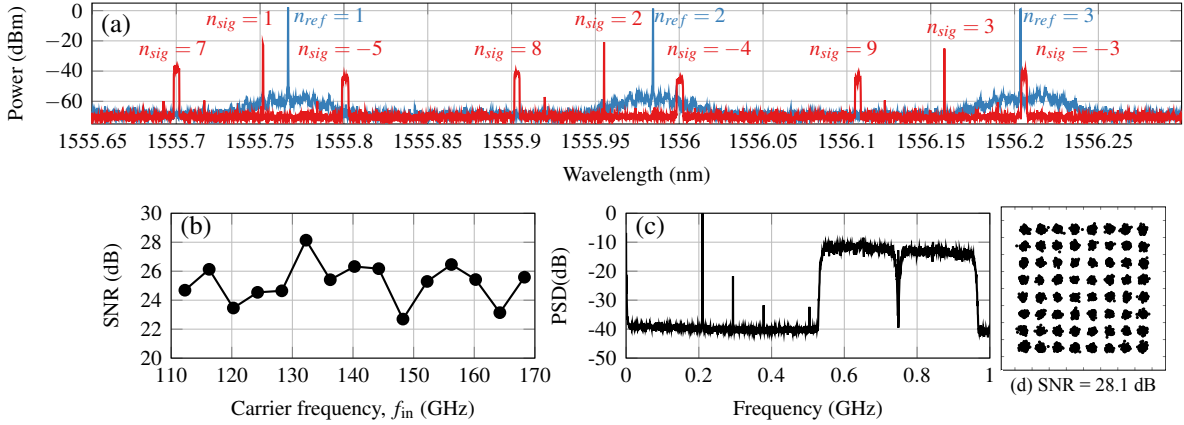

An example of the modulated signal comb is shown in Fig.~\ref{fig:results}(a), showing a portion of the spectrum for reference comb lines $n_\textnormal{ref}=1,2,3$ for an input carrier frequency of 155.82~GHz. Several modulated comb lines can be observed, which are labelled with their originating signal comb line. Also visible are the residual signal comb lines $n_\textnormal{sig}=1,2,3$ due to the limited extinction ratio of the MZM. Most importantly, in the case of $n_\textnormal{ref}= 3$ $n_\textnormal{sig}= -3$ as visible at 1556.2~nm, the modulated comb signal is close to the reference signal (i.e. $<\Delta f/2$) and so will generate a beat signal on the balanced photodector.

Fig.~\ref{fig:results}(b) shows the received signal signal-to-noise ratio (SNR) at different frequencies, across 110 to 170 GHz. The SNR of the received RRC-shaped 64 QAM signals varies from 22.7-28.1~dB with a mean SNR of 25.1~dB. For all frequencies, the normalised generalised mutual information (NGMI) is calculated to be $>0.98$. Part of the SNR variation is due to variations in comb line power, which can be further improved using a more sophisticated pulse-shaping approach~\cite{Torres-Company:08}. Additionally, the unflat gain of the multiplier (about 5 dB variation) in conjunction with amplifier gain limitation causes input power variation that also results in SNR variation. The D-band amplifier has a relatively low $P_{\textnormal{1dB}}$ of 11~dBm that limits the available power to drive the MZM without causing non-linearities. This causes additional modulation loss and creates a tradeoff between non-linearity and OSNR that is typical of microwave photonic techniques~\cite{marpaung2019integrated}. This tradeoff is exacerbated here by the high insertion loss and high V$_\pi$ of the modulator at D-band frequencies. Designing a modulator with more efficient modulation at 110-170~GHz, e.g. through direct coupling of the antenna to the modulator transmission line~\cite{gaier2025wirelessmillimeterwaveelectroopticsfilm}, would allevaite this tradeoff.

An example spectrum for a carrier frequency of 132.252 GHz is shown in Fig.~\ref{fig:results}(c) along with the corresponding recovered constellation diagram in Fig.~\ref{fig:results}(d). The spurs appearing in the recovered spectrum are a result of the residual signal comb lines due to the finite extinction ratio of the MZM. These spurs are deterministic and can be eliminated in DSP. Note that the use of a single balanced detector here requires that the baseband be oversampled by at least a factor of 2 to resolve ambiguities when the downconverted signal overlaps DC and is frequency folded back onto itself, i.e. 4~GSa/s in this case, since $\Delta f/2 = 1$~GHz requires 2~GSa/s. Alternatively, the balanced detector in Fig.~\ref{experiment_setup}(a) could replaced by a 2~GSa/s optical coherent receiver. One limitation of the dual comb technique is that it is not possible to determine the carrier frequency of the detected signal unambiguously without the use of optical filters. This limitation can be solved by a applying a small frequency modulation to one of the comb lines~\cite{o2022architecture}, Alternatively, channel identifying information can be encoded in the transmitted signal for a wireless link, or a bandpass RF filter could be placed before the modulator if the specific application does not require detection across the entire D-band.

\section{Conclusion}

We demonstrate a D-band receiver concept based on dual optical frequency combs, allowing for high resolution detection of narrowband signals across the entire D-band, without the use of frequency tuning or optical filters.

\bibliographystyle{opticajnl}
\bibliography{sample}

\end{document}

%% file: figs/channel_assignment.tex
\begin{tikzpicture}[trim axis left,trim axis right]

\begin{axis} [ylabel= Comb line number, 
              xlabel=Carrier frequency (GHz),
              xmin=110,xmax=170,
              ymax=13,
              ymin=-6,
              height=0.23\linewidth,
              grid=both,
              width=0.4\linewidth,
              cycle list/Dark2,
              ylabel near ticks,
              legend columns =3,
              clip mode=individual,
              log ticks with fixed point,
              legend style={at={(0.5,1.05)}, anchor=south, font=\footnotesize},
              label style={font=\footnotesize},
              tick label style={font=\footnotesize},
              xticklabels={110,120,130,140,150,160,170},
              xtick={110,120,130,140,150,160,170}]

\addplot+ [mark=none, thick] table [x=freq, y=lo, col sep=comma] {data/channel_assignment.csv};
\addlegendentry{Ref};

\addplot+ [mark=none, thick] table [x=freq, y=pos, col sep=comma] {data/channel_assignment.csv};
\addlegendentry{Sig USB};

\addplot+ [mark=none, thick] table [x=freq, y=neg, col sep=comma] {data/channel_assignment.csv};
\addlegendentry{Sig LSB};

\node[anchor=west, align=center] at (axis cs:110,-3) {(b)};


\end{axis}
\end{tikzpicture}

%% file: figs/comb_plot.tex
\begin{tikzpicture}[trim axis left,trim axis right]

\begin{axis} [ylabel= Power (dBm), 
              xlabel=Wavelength (nm),
              xmin=1552.55,xmax=1558.55,
              ymax=20,
              ymin=-75,
              height=0.2\linewidth,
              grid=both,
              width=0.44\linewidth,
              cycle list/Set1,
              ylabel near ticks,
              legend columns =3,
              clip mode=individual,
              log ticks with fixed point,
              legend style={at={(0.5,1.05)}, anchor=south, font=\footnotesize},
              label style={font=\footnotesize},
              tick label style={font=\footnotesize},
                /pgf/number format/1000 sep={},]

\addplot+ [mark=none, thick] table [x=freq, y=psd, col sep=comma] {data/comb_25GHz.csv};
\addlegendentry{$f_{\textnormal{sig}} = 25$~GHz};

\addplot+ [mark=none, thick] table [x=freq, y=psd, col sep=comma] {data/comb_27GHz.csv};
\addlegendentry{$f_{\textnormal{ref}} = 27$~GHz};

\node[anchor=west, align=center] at (axis cs:1552.5,5) {(b)};



\end{axis}




\end{tikzpicture}

%% file: figs/new_results_fig.tex
\begin{tikzpicture}[trim axis left,trim axis right]
\begin{axis} [ylabel= Power (dBm),
                xlabel=Wavelength (nm),
              xmin=1555.65,xmax=1556.3,
              ymax=5,
              ymin=-75,
              height=0.2\linewidth,
              grid=both,
              width=\linewidth,
              cycle list/Set1,
              ylabel near ticks,
              legend columns =3,
              clip mode=individual,
              log ticks with fixed point,
              legend style={at={(0.5,1.05)}, anchor=south, font=\footnotesize},
              label style={font=\footnotesize},
              tick label style={font=\footnotesize},
                /pgf/number format/1000 sep={},]

\pgfplotsset{cycle list shift=1}
\addplot+ [mark=none, thick] table [x=freqref, y=psdref, col sep=comma] {data/modualted_comb_25GHz.csv};
\pgfplotsset{cycle list shift=-1}
\addplot+ [mark=none, thick] table [x=freq, y=psd, col sep=comma] {data/modualted_comb_25GHz.csv};


\node[anchor=west, align=center] at (axis cs:1555.65,-10) {(a) };

\node[color=Set1-A,anchor=south, align=center] at (axis cs:1555.7,-40) {\footnotesize{$n_{sig}= 7$} };
\node[anchor=south, align=center,color=Set1-A] at (axis cs:1555.74,-20) {\footnotesize{$n_{sig}= 1$} };
\node[color=Set1-A,anchor=south, align=center] at (axis cs:1555.8,-40) {\footnotesize{$n_{sig}= -5$} };

\node[color=Set1-A,anchor=south, align=center] at (axis cs:1555.9,-40) {\footnotesize{$n_{sig}= 8$} };
\node[anchor=south, align=center,color=Set1-A] at (axis cs:1555.96,-20) {\footnotesize{$n_{sig}= 2$} };
\node[color=Set1-A,anchor=south, align=center] at (axis cs:1556.02,-40) {\footnotesize{$n_{sig}= -4$} };

\node[color=Set1-A,anchor=south, align=center] at (axis cs:1556.1,-40) {\footnotesize{$n_{sig}= 9$} };

\node[color=Set1-A,anchor=south, align=center] at (axis cs:1556.24,-40) {\footnotesize{$n_{sig}= -3$} };

\node[anchor=west, align=center,color=Set1-B] at (axis cs:1555.765,-5) {\footnotesize{$n_{ref}= 1$} };

\node[anchor=west, align=center,color=Set1-B] at (axis cs:1555.98,-5) {\footnotesize{$n_{ref}= 2$} };
\node[anchor=south, align=center,color=Set1-A] at (axis cs:1556.17,-25) {\footnotesize{$n_{sig}= 3$} };
\node[anchor=west, align=center,color=Set1-B] at (axis cs:1556.2,-5) {\footnotesize{$n_{ref}= 3$} };

\end{axis}
\end{tikzpicture}
\begin{tikzpicture}[trim axis left,trim axis right]

\begin{axis} [ylabel= SNR (dB), 
              xlabel={Carrier frequency, $f_\textnormal{in}$ (GHz)},
              xmin=110,xmax=170,
              ymax=30,
              ymin=20,
              height=0.22\linewidth,
              grid=both,
              width=0.4\linewidth,
              cycle list/Dark2,
              ylabel near ticks,
              legend pos=south west,
              legend columns =2,
              clip mode=individual,
              log ticks with fixed point,
              legend style={font=\footnotesize},
              label style={font=\footnotesize},
              xtick = {110,120,130,140,150,160,170},
              xticklabels = {110,120,130,140,150,160,170},
              tick label style={font=\footnotesize}]

\addplot+ [mark=*, thick,color=black] table [x=FreqGHz, y=SNR, col sep=comma] {data/64QAM_SNR_2.csv};


\node[anchor=west, align=center] at (axis cs:110,29) {(b)};

\end{axis}
\end{tikzpicture}\hspace{1.2cm}\begin{tikzpicture}[trim axis left,trim axis right]

\begin{axis} [ylabel= PSD(dB), 
              xlabel=Frequency (GHz),
              xmin=0,xmax=1,
              ymax=0,
              ymin=-50,
              height=0.22\linewidth,
              grid=both,
              width=0.4\linewidth,
              cycle list/Dark2,
              ylabel near ticks,
              legend columns =3,
              clip mode=individual,
              log ticks with fixed point,
              legend style={at={(0.5,1.05)}, anchor=south, font=\footnotesize},
              label style={font=\footnotesize},
              tick label style={font=\footnotesize},
              yticklabels={-50,-40,-30,-20,-10,0},
              ytick={-50,-40,-30,-20,-10,0}]

\addplot+ [mark=none, thick,color=black] table [x=freq, y=psd, col sep=comma] {data/spectrum_paper.csv};


\node[anchor=west, align=center] at (axis cs:0,-5) {(c)};

\end{axis}
\end{tikzpicture} \includegraphics[width=0.14\linewidth]{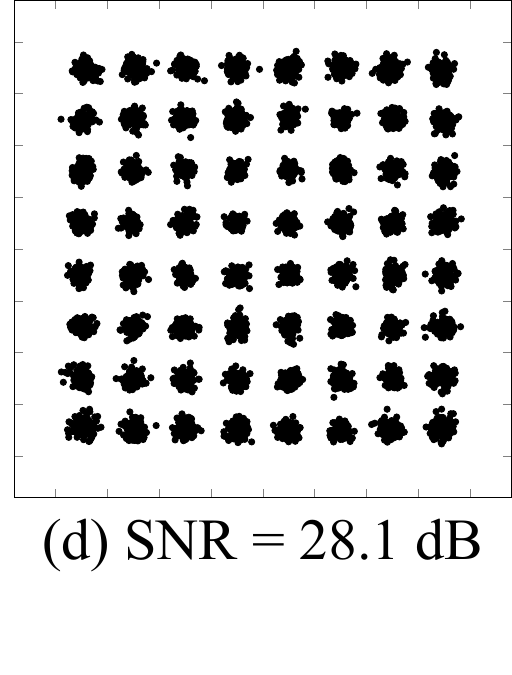}